\documentclass{elsart}
\usepackage{amsmath}
\usepackage{amsfonts}
\usepackage{amssymb}
\usepackage{graphicx}

\begin{document}
\begin{frontmatter}
\title{Properties of Groove Chambers }
\author{I. Reichwein},
\author{ U. Werthenbach \thanksref{mail}},
\author{G. Zech}
\address{Universit\"{a}t Siegen, D-57068 Siegen}
\thanks[mail]{E-mail: werthenb@elfi1.physik.uni-siegen.de}
\begin{abstract}
Groove chambers with different pitch and gap height have been tested. Gas
amplifications of the order of a few times 10$^{3}$ have been obtained.
Combining a groove structure with a GEM pre-amplification foil a gas gain of 10$^{5}$ was
reached. The device is robust and can be produced at low cost in large sizes
by a laser technique.
\end{abstract}
\end{frontmatter}

\section{Introduction}

Since the development of the Microstrip Gas counter \cite{oed88} a large
variety of detectors based on lithographic patterns and gas amplification have
been invented. The expensive and fragile MSGCs with the thin anode lines have
been replaced by simpler and more robust structures. Among the post MSGC
developments (for a review see ref. \cite{sha01}) the GEM \cite{sau97} and
its derivatives the Multi-GEM (MGEM) devices have become the dominant line. A
disadvantage of MGEMs is the requirement of several gas gaps which increase
the mechanical complexity and lead to longer drift times which in turn may
cause overlapping signals at high rate.%

\begin{figure}
[htbp]
\begin{center}
\includegraphics[bb= 10 10 400 300] {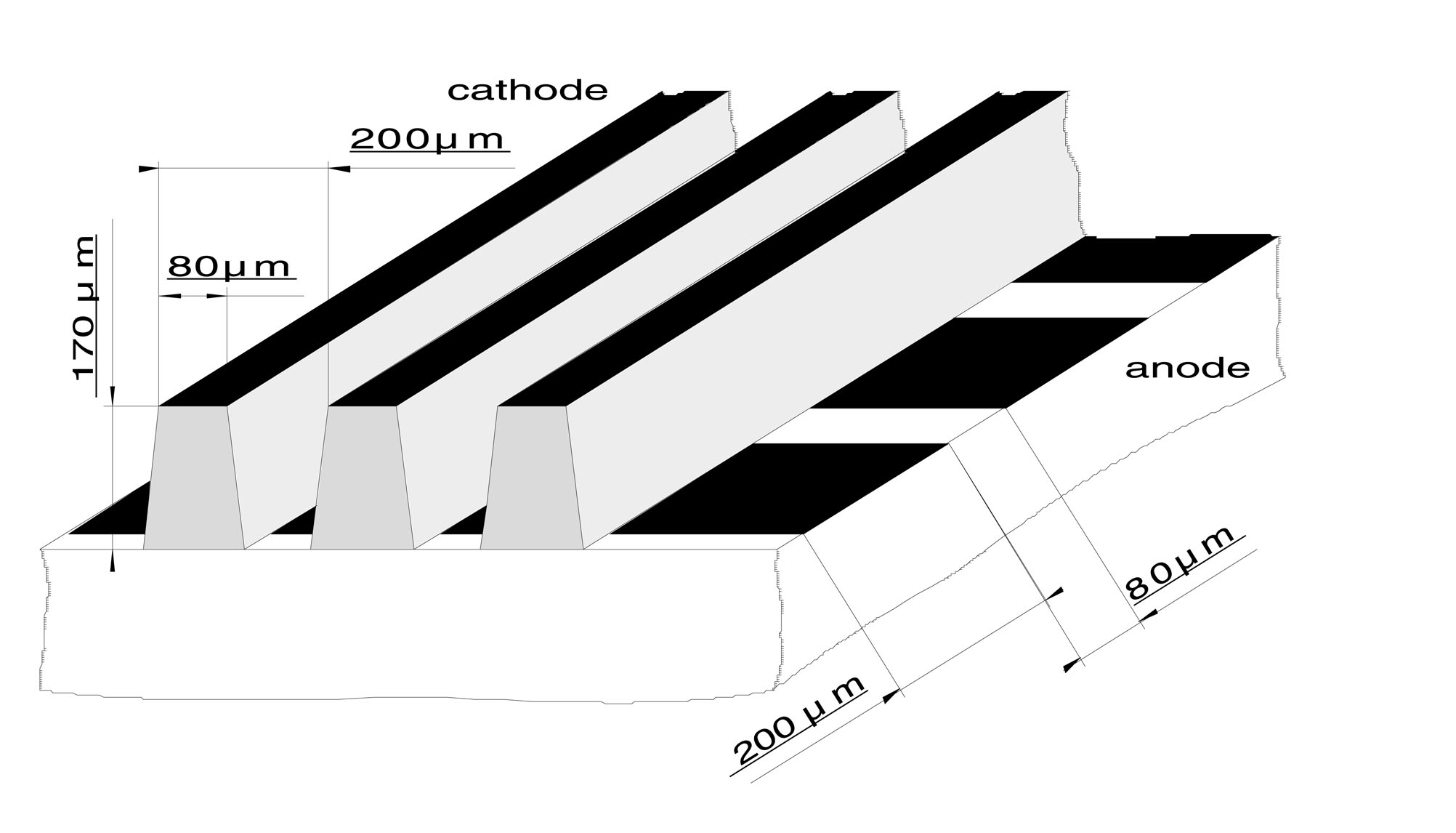}
\caption{Schematic view of a groove chamber with two-dimensional readout.}%
\end{center}
\end{figure}

The groove chamber \cite{bar97} operates in a similar fashion as the GEM.
Combining the readout board with the GEM it suppresses the disturbing last gas
gap of a MGEM and provides in addition a two-dimensional readout. A schematic
picture of a groove chamber is shown in Figure 1.

\section{Construction of the chamber}

The production steps are illustrated in Figure 2. A printed board with the
anode lines, a partially polymerized prepreg of epoxy with aramid fibers and a
copper foil are pressed together. Heating the system to a moderate temperature
the prepreg is fully polymerized and glued to the printed board and the copper
foil. With standard lithographic techniques cathode lines are produced
perpendicular to the anode lines at the top copper layer. In the following
step, a CO$_{2}$ laser removes the prepreg material between the cathode lines.
The copper reflects the laser light and thus screens the spacer material.
Finally the groove has to be cleaned to remove material evaporated during the
laser treatment and deposited at the electrodes or the side walls of the
groove. The desmearing has been performed with a KMnO$_4$ solution
but could also be done by plasma etching.%

\begin{figure}
[htbp]
\begin{center}
\includegraphics[bb= 100 100 400 600]{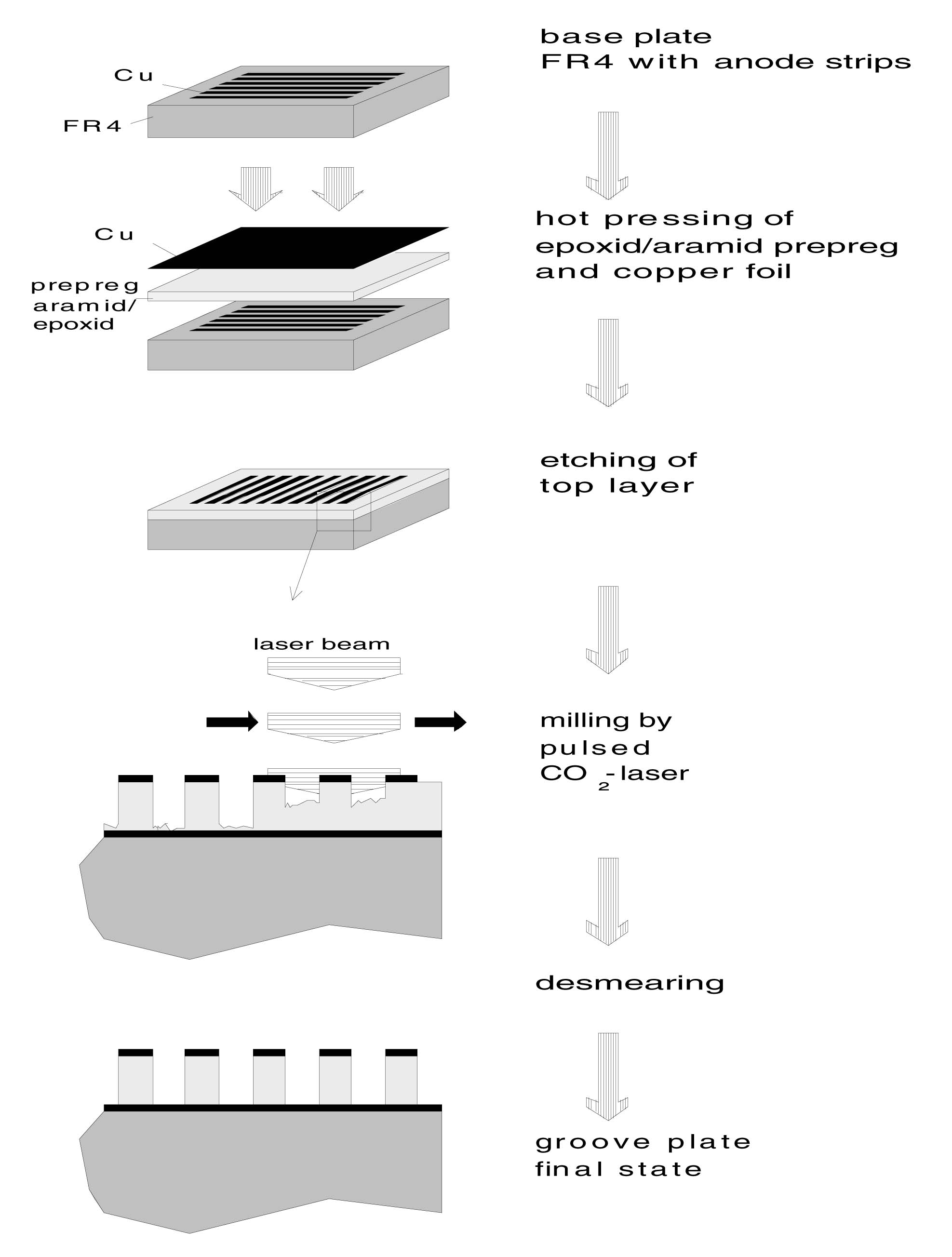}
\caption{Sequence of production steps of a groove structure.}%
\end{center}
\end{figure}

There are no obvious limitations on the size of the detectors from the
production technique. In order to be able to perform a large number of tests
we have produced many small detectors which we cut from plates of 40 cm x 40 cm.%

\begin{figure}
[htbp]
\begin{center}
\includegraphics[bb= 100 100 400 300] {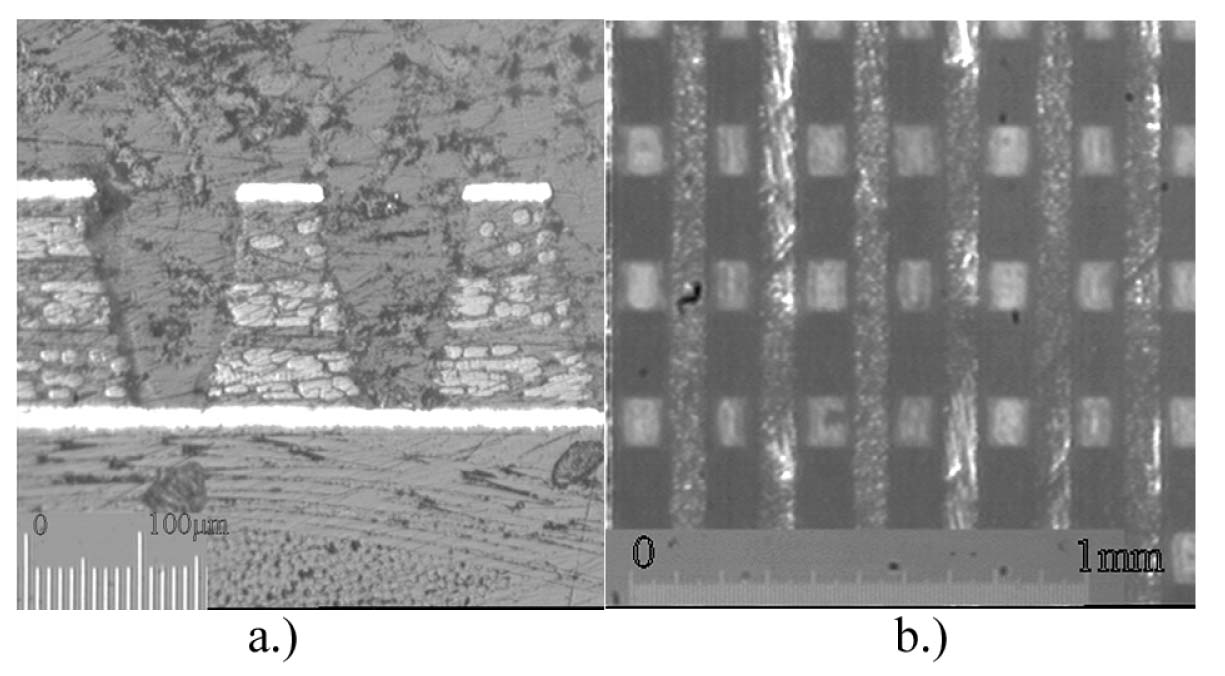}
\caption{Cross section a.) and top viev b.) of a groove structure}%
\end{center}
\end{figure}

Figure 3a shows a cross section of a groove structure\footnote{To produce the
cut through the detector the empty space had been filled with epoxy.}. The
spacers carrying the cathode strips have a trapezoidal shape. The aramid
fibers embedded in the expoxy matrix are well seen. Figure 3b is a top view
showing the regular grid formed by the anode and the cathode lines.

\begin{center}
\begin{table}[htbp]  \centering
\caption{Geometic constants of detectors}
\vspace{5pt}
\begin{tabular}
[c]{|l|l|l|l|}%
\hline
detector & cathode width & pitch & spacer height\\
\hline
\hline
I & 160 & 140 & 220\\ \hline
II & 80 & 120 & 170\\ \hline
III & 70 & 130 & 100 \\ \hline
\end{tabular}
\end{table}
\end{center}

To reduce the number of readout channels we always combined ten cathode strips
and ten anode strips to one readout channel except for ten adjacent anodes
which were readout individually. The sensitive area of the detectors were 31.8
x 28.5 mm$^{2}$. The anode structure is identical in all three chambers. It is 
shown in Figure 1 (anode width: $200 \mu m$, pitch: $80 \mu m$). The three
different cathode configurations which were produced are summarized in Table 1.

\begin{figure}
[htbp]
\begin{center}
\includegraphics[bb= 100 100 400 300] {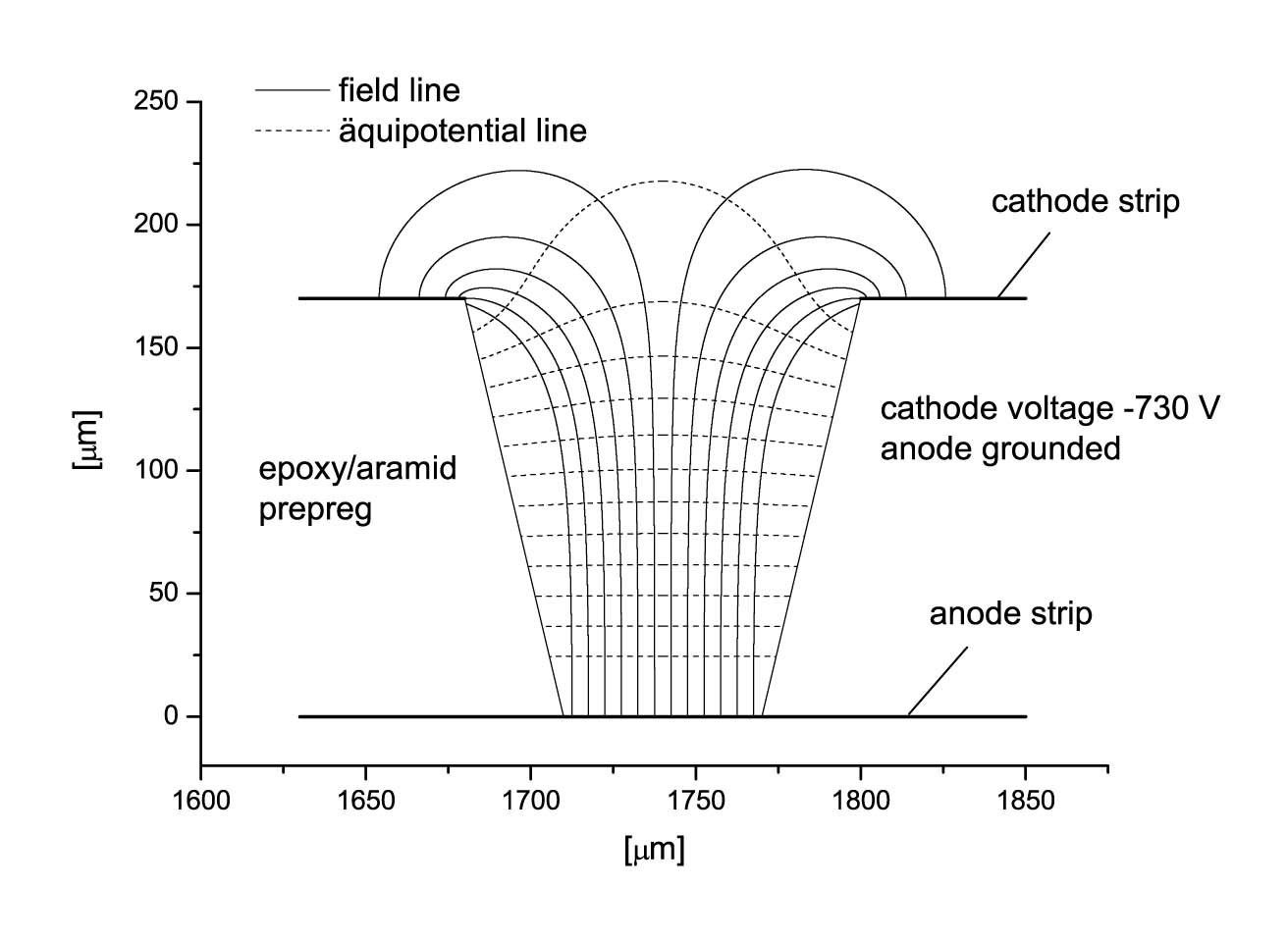}
\caption{Electric field of a groove chamber.}%
\end{center}
\end{figure}

We used a standard gas mixture of 82\% CO and 18\% Ar. The gas gap was 8 mm
and the drift field was about \ 2 kV/cm. Figure 4 shows the result of a field
simulation. The field is rather insensitive to the dielectric constant of the spacer
material. In reality, it will be distorted by positive ions deposited at
the side walls of the groove.

We used standard readout amplifiers with a shaping time of $\approx20 ns$
and identical timing for the anode and the cathode channels.

The detector II was operated with GEM pre-amplification. The GEM was
produced at the CERN workshop and consisted of a 50 $\mu$m thick polyimid foil
cladded on both sides by 10 $\mu$m thick copper layers. The holes were
arranged in form of a hexagonal grid, the distance between neighboring holes
being 125 $\mu m$. Hole diameters were 40 $\mu m$ at the centre and 85 $\mu m
$ at the copper sides. The foil was stretched 3 mm above the groove structure
and 5.5 mm below the drift electrode. The GEM was operated at a maximum
voltage of 440 V, corresponding to a gas amplification of about 40. Drift and
transfer fields were around 2 kV/cm and 2.2 kV/cm, respectively.%

\begin{figure}
[htbp]
\begin{center}
\includegraphics[bb= 100 100 400 300] {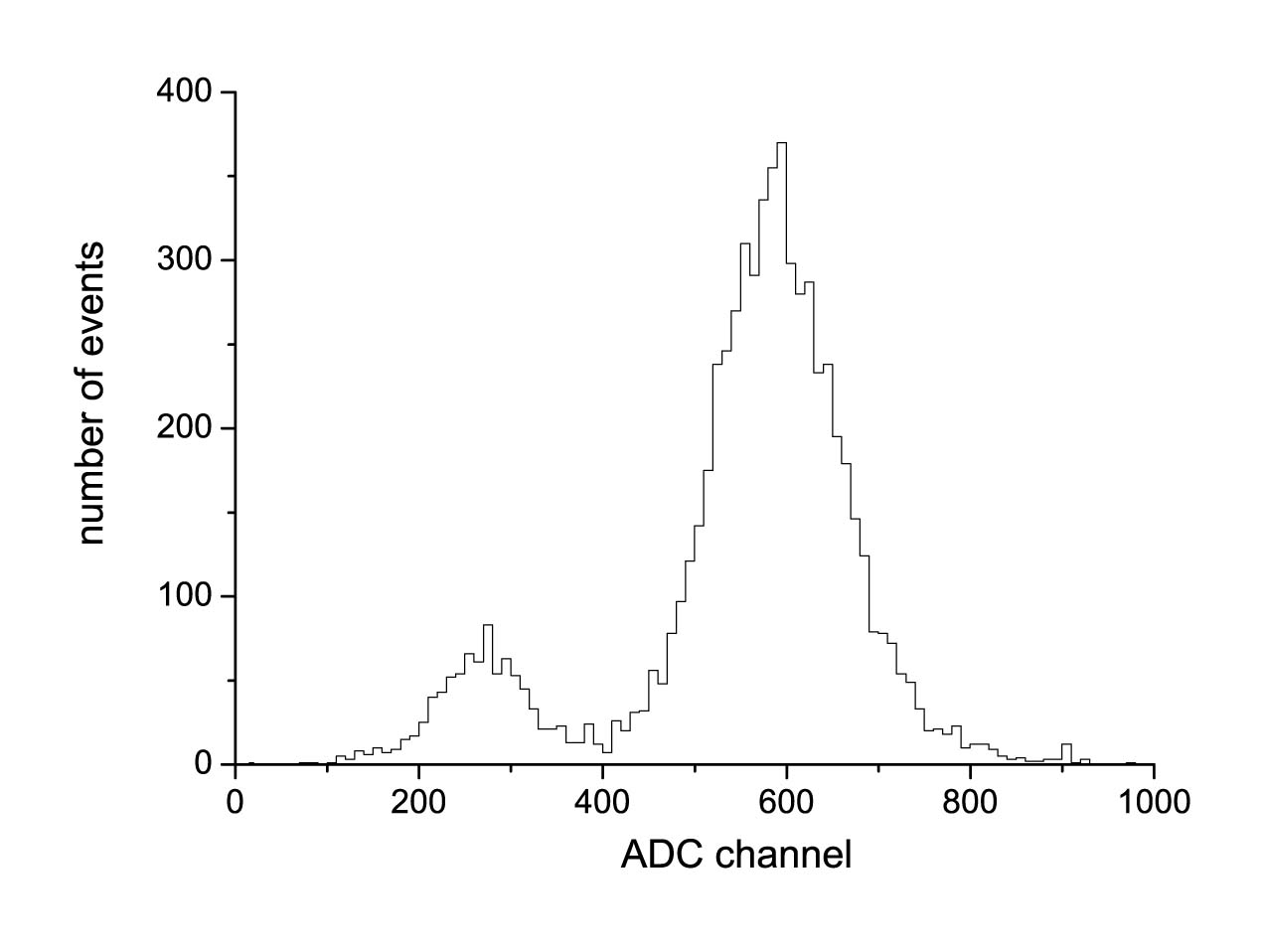}
\caption{Anode pulse height distribution of an Fe-55 source.}%
\end{center}
\end{figure}

\section{Results}

Figure 5 shows Fe-55 spectra with an energy resolution of about 27$\%$, very similar to that of
MSGCs or GEMs. The gas amplification is very sensitive to the gap height. Thus
we expected rather large gain variations across the chambers. Even though our
chambers were quite small, the observed variation reached almost 20\%.

The cathode signals are very similar to the anode signals and show a similar
energy resolution. Figure 6 is a scatter plot of anode pulse height versus
cathode pulse height. It shows the expected strong correlation which could be
a useful tool in the analysis of overlapping signals.%

\begin{figure}
[htbp]
\begin{center}
\includegraphics[bb= 100 100 400 300] {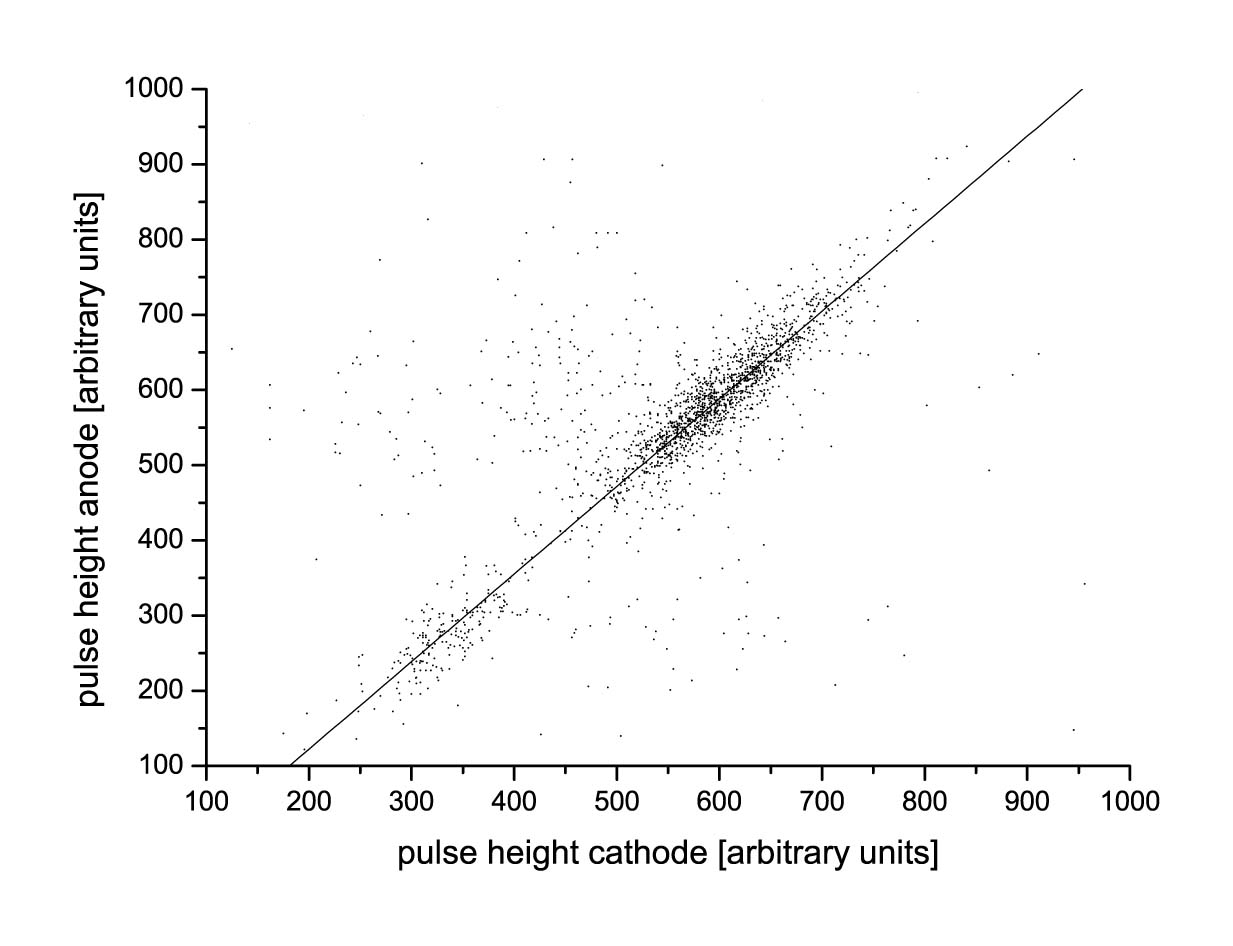}
\caption{Scatter plot of pulse heights derived from anodes and cathodes,
respectively.}%
\end{center}
\end{figure}

The average anode multiplicity (anodes fired per primary photon) is about 2.2
for a threshold of \ 1\% of the total energy. In the pulse height spectrum for
the 18 \% of cases where three anodes fire the escape peak is almost
completely absent indicating that a large fraction of these events is due to
an escape photon converting at a small distance from its origin.
Correspondingly, the escape peak is very pronounced in the 9\% of cases where
only a single anode shows a signal.

The cathode multiplicity must be at least
two. Experimentally, it can be deduced only indirectly because of the cathode
grouping. We find a mean value of 2.7$\pm$0.7.%

\begin{figure}
[htbp]
\begin{center}
\includegraphics[bb= 100 100 400 300] {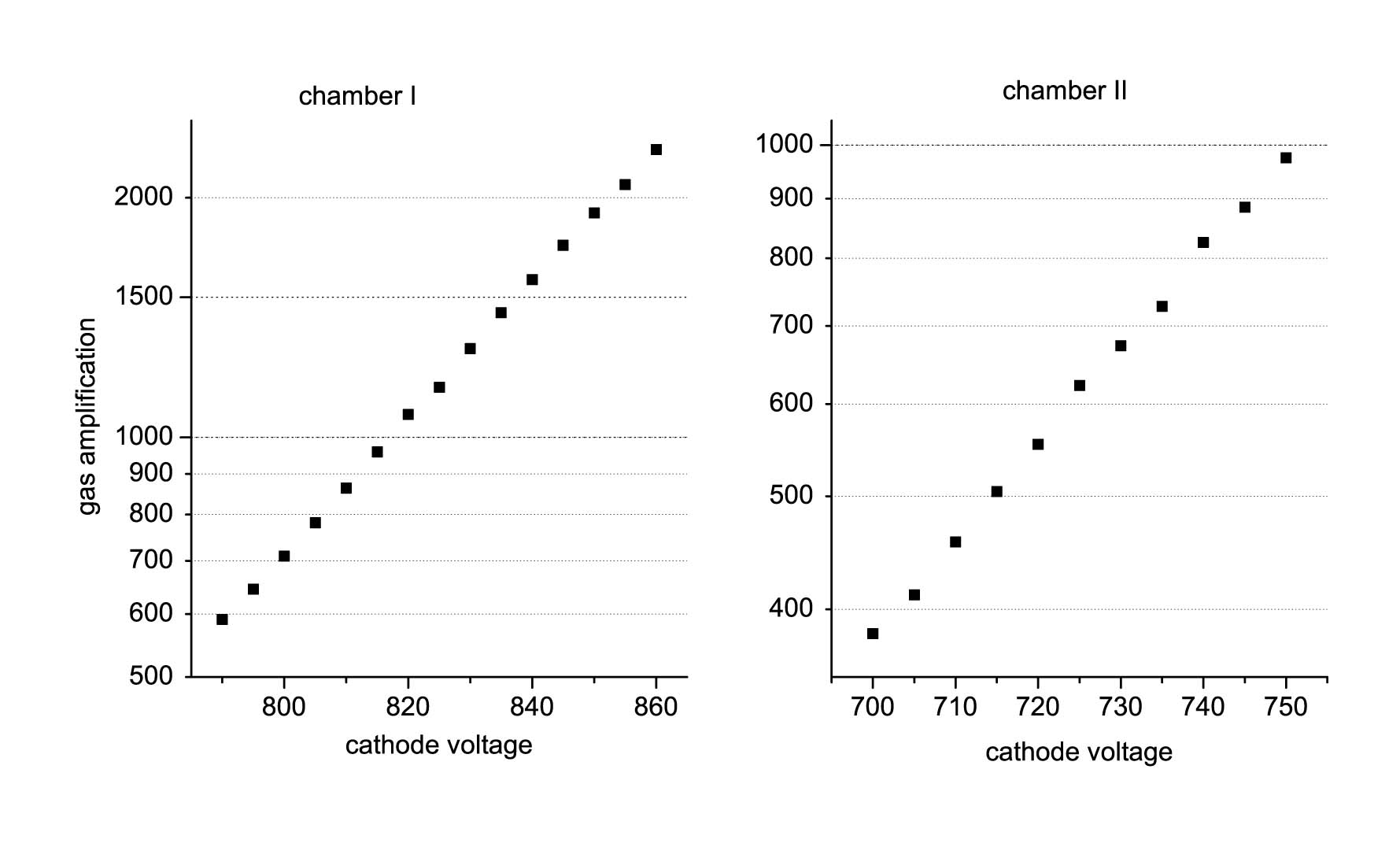}
\caption{Gas amplification for two different groove chambers.}%
\end{center}
\end{figure}

We have gauged the gas amplification using the chamber with GEM. Raising the
GEM voltage from zero to 440 V at fixed groove voltage we could determine the
GEM amplification and the ADC channel-to-amplitude relation.

Figure 7 compares the gas
amplifications of chambers I and II. The discharge limits occurred at a gain
of 2000 for chamber I and at 1000 for chamber II. The chambers are not
affected by spurious discharges, however continuous sparking of course
prevent measurements. Since the spacer of chamber I is much higher than that
of chamber II, naively, we could expect a much higher gain for the former.
In fact only a moderate improvement factor of two was achieved. The discharge
limit is most likely given by the non-uniformity of the electric field caused
by surface currents at the groove walls.%

\begin{figure}
[htbp]
\begin{center}
\includegraphics[bb= 100 100 400 300] {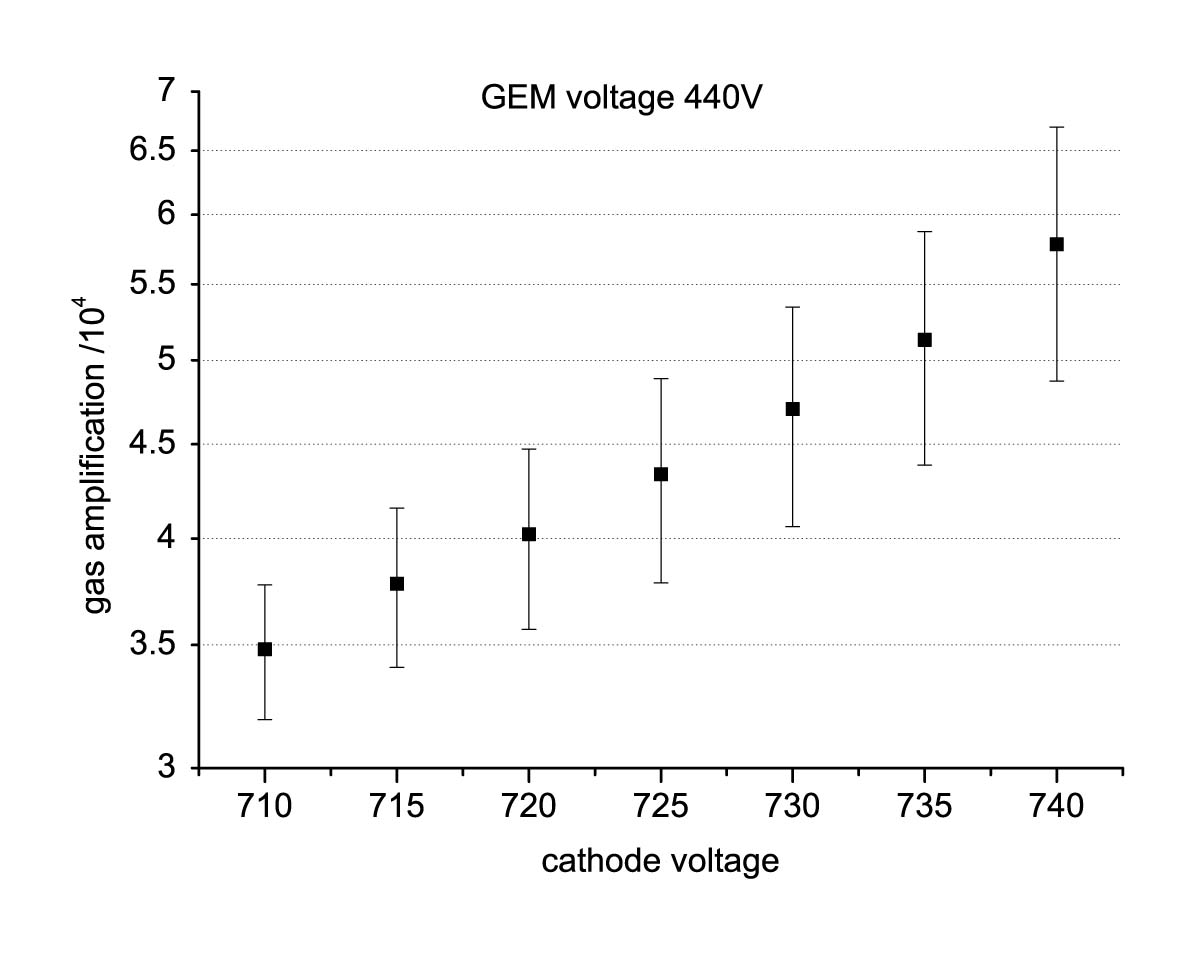}
\caption{Gas amplification of a groove chamber with GEM pre-amplification.}%
\end{center}
\end{figure}

With the groove chamber with the GEM pre-amplification, gas amplifications in
excess of 5$\times10^{4}$ were reached. At this regime no sizable deviation of
the exponential increase of the gain with the applied voltages was observed
(see Figure 8). The relatively large errors in Figure 8 steam from the
combination of the various gain factors and the necessity to use voltage
dividers to access the full amplification range with a single amplifier.

Results from a similar device to ours have been presented in ref. \cite{bel99}.

\section{Possible improvements and conclusions}

We are convinced that the present limitations of the groove chamber are
related to the ill defined electric properties of the plastic spacer which
does not lead to an uniform potential drop between cathode and anode. Ideally,
the material should have electronic conductivity independent of the
irradiation. Since large currents have to be avoided for large detectors a
large surface conductivity combined with a low bulk conductivity were
desirable. Coating the insulator with polycrystallic carbon which was
successful for MSGCs and partially also for GEMs \cite{bei99} is not possible
because the CVD process is hampered by the metal electrodes.

We have investigated different prepreg materials and bakelite which has a
rather good conductivity. We also tried the famous linseed oil coating which
successfully reduces discharges in resistive plate chambers. All these efforts
were unsuccessful. Nevertheless we are convinced that there exist
technological solutions which would solve the problem and provide gas
amplifications of grove chambers of 10$^{5}$. The necessary developments
however are expensive as long as there is no interest and engagement from industry.

The results obtained so far indicate that large detectors with an additional
single GEM could be a solution superior to triple GEM detectors in many
applications. It is mechanically and electronically much simpler. To
demonstrate this, large scale detectors have to be produced and tested in
hadron beams.

\textbf{Acknowledgement}

We are very grateful to Mr. Lippert and Mr. Simson at Nelco Dielektra
for producing the structures  and to Mr. B\"{u}chner at Schweizer Elektronik AG
for performing the desmearing.

\bigskip
\end{document}